\documentstyle[prl,aps,epsfig,preprint]{revtex}

\begin{document}
\draft
\def\be{\begin{equation}}
\def\ee{\end{equation}} 
\def\bfi{\begin{figure}}
\def\efi{\end{figure}}
\def\bea{\begin{eqnarray}}
\def\eea{\end{eqnarray}}
\title{Universality of the off-equilibrium response function
in the kinetic Ising chain}

\author{Federico Corberi$^{1,\dag}$, Claudio Castellano$^{2,*}$, 
Eugenio Lippiello$^{1,\ddag}$ and Marco Zannetti$^{1,\S}$}
\address{
$^1$ Istituto Nazionale per la Fisica della Materia,
Unit\`a di Salerno and Dipartimento di Fisica, Universit\`a di Salerno,
84081 Baronissi (Salerno), Italy \\
$^2$ Istituto Nazionale per la Fisica della Materia, Unit\'a
di Roma ``La Sapienza'', Piazzale Aldo Moro 2, 00185 Roma, Italy}

\maketitle
 
\begin{abstract}
The off-equilibrium response function $\chi (t,t_w)$ and autocorrelation
function $C(t,t_w)$ of an Ising chain with spin-exchange dynamics 
are studied numerically and compared with the same quantities in 
the case of spin-flip dynamics. It is found that, 
even though these quantities
are separately different in the two cases,
the parametric plot of $\chi (t,t_w)$ versus $C(t,t_w)$ is the same.
While this result could be expected in higher dimensionality,
where $\chi (C)$ is related to the equilibrium state, it is far from
trivial in the one dimensional case where this relation does
not hold. The origin of the universality of $\chi(C)$ is traced back to the
optimization of domains position with respect to the perturbing
external field.
This mechanism is investigated resorting to models with a single domain moving
in a random environment.

\end{abstract}

\section{Introduction} \label{intro}

The generalization of the fluctuation-dissipation theorem
(FDT) to slowly relaxing systems, such as glasses,
is an issue of foremost importance for understanding
non-equilibrium processes. 
Ordinary FDT relates the autocorrelation $C(t,t_w)$
and the integrated response $\chi(t,t_w)$ functions, 
which both depend in equilibrium 
on the time difference $t-t_w$, where $t_w$ is the time
elapsed after the sample preparation.
A general feature of slow kinetics, instead,
is the aging property, namely the dependence of the time-scale
of relaxation on the time $t_w$.
This feature generally shows up both in $C(t,t_w)$ and $\chi(t,t_w)$.
In the context of mean-field models
for spin glasses it was shown~\cite{cugliandolo93} that, for
large $t_w$, $\chi (t,t_w)$
depends on the two times through the autocorrelation function alone
\be
\chi (t,t_w)=\chi [C(t,t_w)].
\label{cug}
\ee
This property holds quite generally in
a wide class of aging systems where
deviations from the ordinary equilibrium FDT, namely 
$T\chi (C)=C(t,t)-C(t,t_w)$, 
result in a nontrivial fluctuation dissipation ratio 
$X(C)=-Td\chi (C)/dC$.
Recently, a theorem has been proven~\cite{franz98}
linking $X(C)$ to static properties
\be
\left .\frac {dX(C)}{dC}\right |_{C=q}=P(q)
\label{statdin}
\ee
where $P(q)$ is the equilibrium probability distribution of the overlaps.
This opens the way to a classification of aging systems according to
the structure of their equilibrium states~\cite{parisi99} and to the
recognition that the properties of the response to a perturbation 
are universal in systems sharing the same overlap distribution.

The ordering process of ferromagnetic systems 
provides a simplified framework for the study of the off-equilibrium
FDT because the main features of slow relaxation are fully
exhibited but the structure of their equilibrium state 
is simple and exactly known. 
While~(\ref{cug}) is generally obeyed~\cite{parisi99,barrat98,corberi2001},
the validity of~(\ref{statdin}) in this case depends on dimensionality.
For $d>1$, (\ref{statdin}) holds asymptotically, implying the same 
fluctuation dissipation ratio, 
and hence of $\chi (C)$, for all these systems~\cite{parisi99}. 
This applies, in particular, to different dynamical realizations
of the same Hamiltonian model
such as ferromagnets with non-conserved
(NCOP) or conserved (COP) order parameter~\cite{barrat98}. 
However the picture is totally different in the case of the Ising chain.
With NCOP,~(\ref{statdin}) is not obeyed~\cite{corberi2001}
and, instead, a non trivial $X(C)$ is found~\cite{godreche2000}  
that cannot be connected to static properties.

These results suggest that the nature of $X(C)$ 
is essentially dynamical in this case.
A natural question, then, is about which properties of the kinetics
are reflected by $\chi (C)$. In order to address this point,
we study in this paper the response of the one-dimensional
Ising model with Kawasaki spin-exchange dynamics quenched to a
low temperature, in the scaling regime preceding equilibration. 
In this case the order parameter is conserved and the microscopic mechanism 
whereby coarsening of domains is produced differs completely from
the dynamics with single spin flip. With NCOP interfaces are independent
brownian walkers whose density is progressively reduced
due to annihilation events. With COP, instead,
the motion of interfaces is mediated by evaporation, diffusion and
recondensation of single monomers.
Despite this completely different character of the dynamics,
we show that $\chi (C)$ is the same for COP and NCOP.
Due to the violation of the hypotheses of theorem~(\ref{statdin}) this  
universal character cannot be traced back to statics 
but it is more likely to have a common dynamical origin.
Since the basic coarsening mechanisms with COP or NCOP are profoundly
different, other kinetic properties, of a more general
and fundamental nature, determine $\chi (C)$ . 
The analysis carried out in this article shows that 
the total response can be viewed as due to the 
elementary contributions given by single domains.
In the kinetic process domains coarsen and translate
and the complex interplay between these two mechanisms
produces the response.
Starting from this idea we introduce simple models where a
single domain is allowed to diffuse in a random environment.
In this framework, the elementary response generated by the domain
can be studied and from its knowledge the behavior of the
original Ising chain is inferred by means of scaling arguments.
The details of the rules for the motion 
do not change the overall behavior of the response.
This approach provides a clear physical interpretation of how the
response is produced in the Ising model.

These considerations are of a general nature and apply, in principle,
to any dimension. On the other hand, 
the occurrence in $d=1$ of an  off-equilibrium response which never
vanishes, as opposed to the cases with $d>1$, is due to the special
character of domain walls motion in one dimension.
Actually, in $d>1$ the evolution of a domain 
is the result of two competing drives.
The first is the tendency to lower surface tension by making interfaces
straight. The second is the drift towards regions where the random field
is favorable.
For long times the first mechanism always prevails, and the response generated
by the drift of domains is negligible~\cite{corberi2001}.
In $d=1$, instead, domain walls are point-like and surface tension does
not play any role; moreover the drift mechanism is so efficient as to
generate a non-vanishing response even in the limit of large times, when
the interface density decreases to zero.

This paper is organized in six Sections. Section~\ref{unpe}
is devoted to a description of the COP dynamics of the unperturbed
Ising chain. In Section~\ref{pert} the effects of a perturbation
are discussed and the response function is introduced, showing
the analogy with the NCOP dynamics and the universality of $\chi (C)$.
Models for a single diffusing domain are discussed in 
Sections~\ref{onerigid} and~\ref{twokawa},
where scaling arguments are presented 
to illustrate the common origin of the response within the two types of
dynamics. In Section~\ref{conclu} we discuss the relevance
of our results for different systems and draw some conclusions. 
 
\section{Unperturbed dynamics} \label{unpe}

We consider the one-dimensional Ising model with ferromagnetic
nearest-neighbor coupling constant, whose Hamiltonian is
\be
{\cal H}_0(\{s_i \})=-J\sum _{i=1} ^N s_i s_{i+1},
\ee
where $s_i=\pm 1$. The system is quenched from an uncorrelated 
high temperature equilibrium state to the final temperature $T$.
Evolution takes place through Kawasaki spin-exchange dynamics, i.e.
swaps between antiparallel nearest-neighbor spins. 
In this way magnetization is a conserved quantity. The model 
describes lattice gases or binary alloys.    
The probability
of exchanging $s_i$,$s_{i+1}$ is assumed to be
\be
p_T=\min \left [ e^{-\frac {\Delta E}{T}},1\right ]
\label{flipprob}
\ee
where $T$ is measured in units of the Boltzmann constant and 
$\Delta E =2J(s_{i-1}s_i+s_{i+1}s_{i+2})$ is the energy change.

The basic features of the dynamics following an instantaneous quench
are discussed in~\cite{cornell91}. Depending on the energy change  
elementary moves can be distinguished into three classes with
$\Delta E=4J,0,-4J$. The first kind of process is evaporation,
namely the separation of a spin from the boundary of a domain.
Processes with $\Delta E=0$ are the diffusion
of a single spin (monomer) in the bulk of a domain of the other phase.
When a diffusing monomer reaches an interface a condensation event occurs: 
The spin joins a domain.
This process implies an energy change $\Delta E=-4J$. 

\begin{figure}
\centerline{\psfig{figure=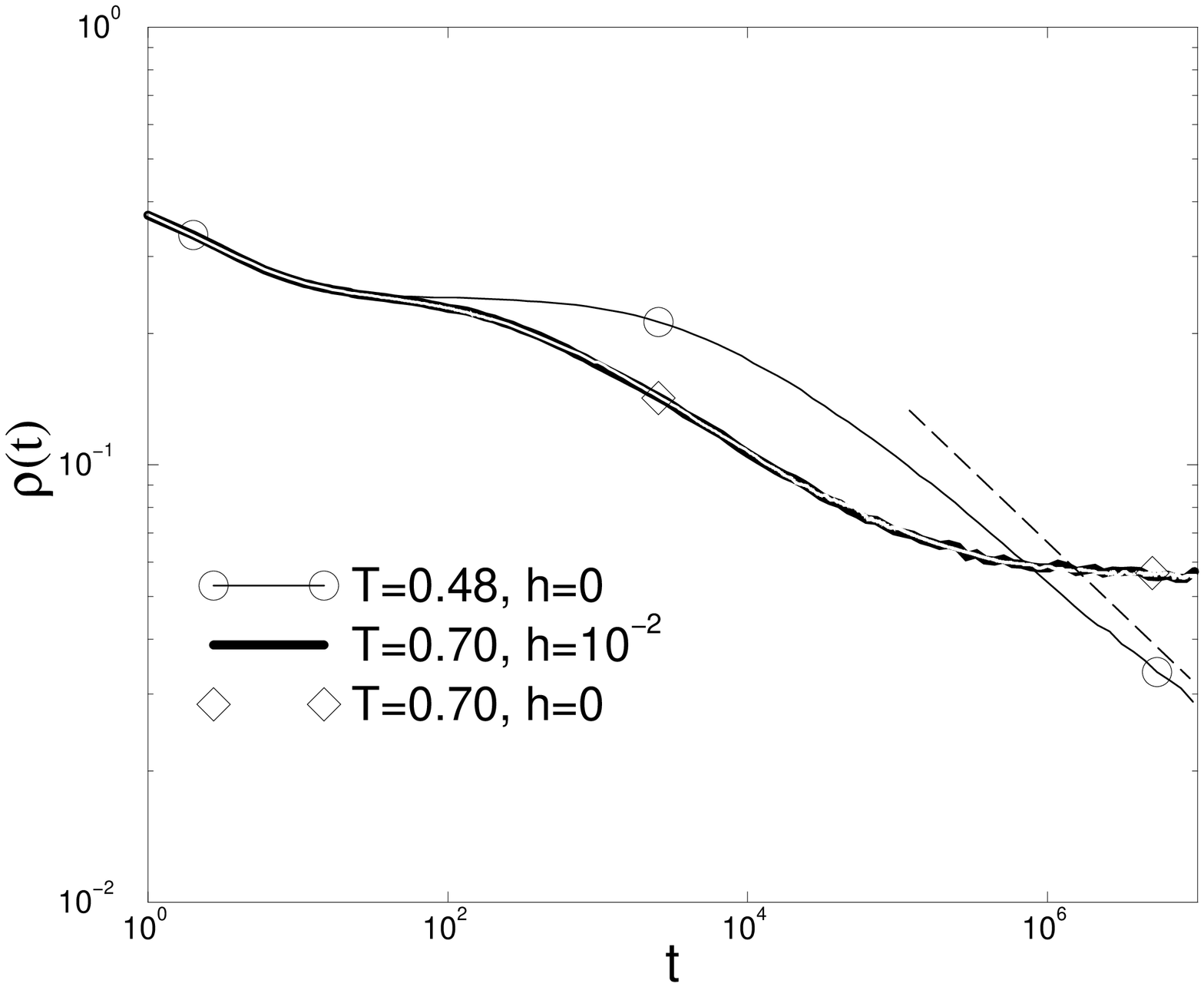,width=9cm,angle=-90}}
\caption{The kink density $\rho $ is plotted against time 
for quenches with $T=0.48$ and $h=0$ (black solid line) and 
$T=0.7$, both with the perturbation ($h=10^{-2}$, black bold solid line)
and without ($h=0$, white solid line collapsing on the
curve with $h=10^{-2}$).
Data refer to numerical simulations of a system of $N=10^5$
spins, averaged over 10 realizations.
The dashed line represents the $t^{-1/3}$ law.}
\label{rho}
\end{figure}

Evaporation is an activated process occurring over
a characteristic time $\tau _{ev}=\exp (4J/T)$.
For low temperatures $\tau _{ev}$ is large and one observes a long interval
$t\ll\tau _{ev}$ during which evaporation practically does not happen.
In this regime a reduction of the kink density $\rho(t)$
can be obtained only by the diffusion and
condensation of the monomers present in the initial state.
In order to do this single spins move a distance of the order of the initial 
coherence length $\xi $ in a typical time $\xi ^2$.  
This leads to a decay~\cite{cornell91,privman92} 
of the density of diffusing single spins over a characteristic
time $\tau _s\sim \xi ^2$. In the case we are concerned with, 
a quench from very high temperature, $\tau _s$ is of order unity 
and one observes a fast decay to a plateau on short time-scales
(Fig.~\ref{rho}).
In the regime $\tau _s \ll t \ll \tau _{ev}$ no diffusing
spins are left and $\rho (t)$ remains constant.
At times of order $\tau _{ev}$ evaporation events begin to occur and
the dynamics {\it restarts} (Fig.~\ref{rho}).
Evaporated spins diffuse and they may recondense on a kink different
from the one where they were emitted.
This is the well known mechanism leading to the decay
$\rho (t)\propto (t/\tau _{ev})^{-1/3}$\cite{cornell91}:
in this regime, dynamic scaling is obeyed.
This behavior lasts until $t\simeq \tau_{eq}^{COP}$ such that
$\rho (\tau _{eq}^{COP})= \rho _{eq}\simeq \exp {(-2J/T)}$,
the equilibrium kink density.
At this time domains reach a size such that a second monomer is
emitted when the first one is still diffusing. When they meet they form
a stable dimer and this process exactly balances the domain 
annihilation due to the 
evaporation-condensation mechanism, so that $\rho$ keeps its equilibrium
value. 
Regarding the data presented in Fig.~\ref{rho}, for $T=0.7$ 
the exponent $1/3$ is not clearly observed because the system
equilibrates too soon. For $T=0.48$ the effective exponent
gradually decreases toward $1/3$.
At the longest times simulated the effective exponent is $\simeq 0.3$.

\begin{figure}
\centerline{\psfig{figure=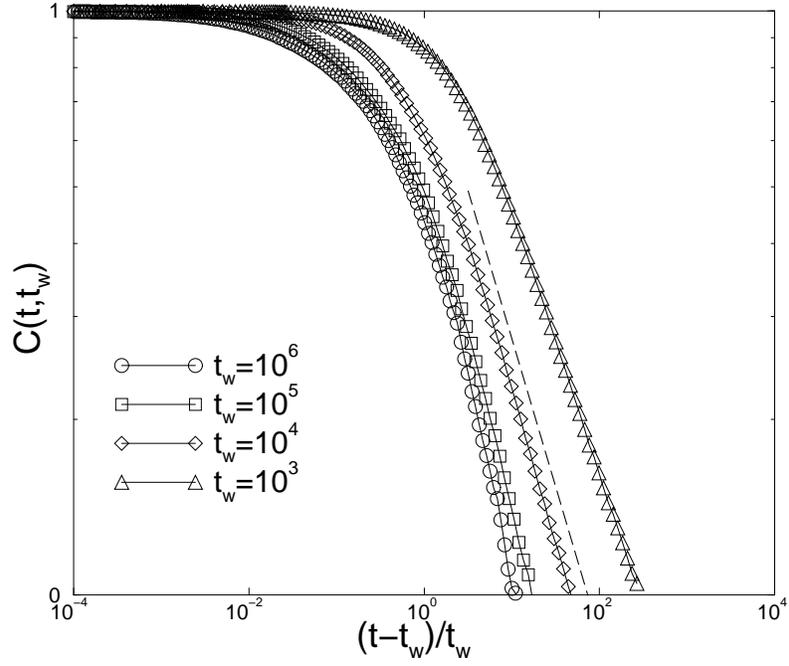,width=9cm,angle=-90}}
\caption{The autocorrelation function $C(t,t_w)$ is plotted against
rescaled time $(t-t_w)/t_w$. The dashed line represents the law
$[(t-t_w)/t_w]^{-1/2}$.}
\label{auto}
\end{figure}

The autocorrelation function is defined as 
\be
C(t,t_w)=\frac {1}{N} \sum _{i=1}^N \langle s_i(t)s_i(t_w) \rangle,
\label{c}
\ee
where $\langle \cdots \rangle$ indicates thermal averaging.
This quantity is shown in Fig.~\ref{auto}. 
$C(t,t_w)$ strongly depends on the range of times
$(t_w$,$t)$ considered. In the case $t_w\ll\tau _{eq}^{COP}$ considered here
the behavior of the autocorrelation function is different for
$t \ll \tau_{eq}^{COP}$ or $t \gg \tau _{eq}^{COP}$.
For $t \ll \tau _{eq}^{COP}$ $C(t,t_w)$ 
decays as $[\rho (t)/\rho (t_w)]^\lambda$.
The exponent $\lambda $ depends on $t_w$ as follows:
When $t_w=0$, the upper bound $\lambda \le d$, originally
proposed by Fisher and Huse~\cite{Fisher1988} provides the correct
value $\lambda =1$ for the conserved $d=1$ Ising model, as shown
analytically and numerically in~\cite{Majumdar1994}.
Instead, for $t_w$ chosen well inside the scaling regime, 
Yeung, Rao and Desai~\cite{Yeung1996} found a lower bound
$\lambda \ge 3/2$ for $d=1$. 
Since $\lambda =1$ for $t_w=0$ this constraint implies 
the dependence of $\lambda $ on $t_w$. 
To our knowledge, there are no
results for the actual value of this exponent when $t_w\neq 0$.
From the data presented in
Fig.~\ref{auto} one observes a power-law decay consistent with 
$C(t,t_w)\sim (t_w/t)^{1/2}$ for the cases $t_w=10^4,10^5$.
Recalling that $\rho (t)\sim t^{-0.3}$ in the range of time considered, 
one obtains $\lambda \simeq 1.66 >3/2$.
In this way we show that the
lower bound determined in~\cite{Yeung1996} is correct and that the value 
of $\lambda$ with $t_w$ chosen inside the scaling
regime is different from the case with $t_w=0$.
Actually, the value $\lambda \simeq 1.66 $ may indicate that the
value $\lambda =3/2 $ could be asymptotically correct. 
In order to check this point 
lower temperatures and larger waiting times should be considered.  
By plotting $C(t,t_w)$ against $\rho (t)/\rho (t_w)$ we have checked
that also the curve with $t_w=10^3$ gives the same exponent 
$\lambda >3/2$, whereas the smaller exponent observed in Fig.~\ref{auto} is
simply due to $\rho (t)$ decaying with an effective
exponent considerably smaller than $1/3$ in the range of times plotted
in the Figure.
For $t_w=10^6$ the curve starts decaying with the same exponent
$\lambda =3/2$ but then the decrease becomes faster,
indicating that the system is close to reaching equilibrium.
A second important observation, regarding Figure~\ref{auto}, 
is the convergence of $C(t,t_w)$ towards the scaling behavior
$C(t,t_w)=\hat C(t_w/t)$, as expected quite generally for slow
relaxation~\cite{Bouchaud1997} and in particular for coarsening
systems~\cite{Bray1994}. However, differently from NCOP,
the convergence in this case is very slow and very large
$t_w$ must be considered in order to exhibit a good data-collapse. 

\section{Response to a perturbation} \label{pert}

Let us consider the Ising model quenched to temperature $T$ in zero field.
At time $t_w$ a random field 
\be
h_i=h\epsilon _i
\label{campo}
\ee
is applied, so that the hamiltonian is changed into
\be
{\cal H}={\cal H}_0-\sum _{i=1}^N h_i s_i.
\label{ham}
\ee
The field takes randomly only two values, $\epsilon _i =\pm 1$,
with expectations
\be
\overline {\epsilon _i} =0
\label{meanh}
\ee
\be
\overline {\epsilon _i \epsilon _j } = \delta _{i,j}.
\label{variah}
\ee
The probability of exchanging two spins $s_i,s_{i+1}$ is~(\ref{flipprob})
with $\Delta E =2J(s_{i-1}s_i+s_{i+1}s_{i+2})+2h(s_i \epsilon _i
+s_{i+1}\epsilon _{i+1})$.
We consider $h/T$ sufficiently small in order to be in the linear 
response regime.
We are interested in the scaling regime, i. e. times such that
$\rho(t) \gg \rho_{eq}$.
Moreover we want the qualitative features of the dynamics, presented above,
to be unaffected by the external field. This imposes an additional
constraint $\rho(t) \gg \xi^{-1}(h)$, where $\xi(h) = 4J^2/h^2$ is the
Imry-Ma length~\cite{Imry75}.
For the values of $h$ and $T$ considered, $\xi^{-1}(h) \ll \rho_{eq}$
so that $\rho (t)$ is unchanged by the presence of the random
field from the instant of the quench up to equilibration (Fig.~\ref{rho}).

We consider the integrated response function 
\be
\chi (t,t_w)=
\lim _{\frac{h}{T}\to 0}\frac {1}{Nh} \sum _{i=1}^N \overline
{\epsilon_i \langle s_i\rangle _h},
\label{resp}
\ee
where $\langle \cdots \rangle _h$ denotes average in presence of the
external field.
Before discussing the response of the model with COP, let us
briefly recall the behavior with NCOP.
In this case equilibrium is reached because spins are flipped spontaneously
in the bulk of ordered domains; the characteristic time for this process
is $\tau_{eq}^{NCOP} = \exp(4J/T)$.
The asymptotic value of the response function is the equilibrium
susceptibility 
\be
\chi _{eq}=1/T,
\label{equi}
\ee
namely  $\lim _{t\to \infty }T\chi (t,t_w)=1$.
In equilibrium  the bulk of domains is responsible for the response
because spins anti-aligned with the random field are more likely
to be reversed by thermal excitations.
However, for times $t<\tau _{eq}^{NCOP}$ spins in the bulk do not flip;
the response observed in this regime 
is then of non-equilibrium nature.
Since the bulk is frozen, $\chi (t,t_w)$ 
must be provided by interfaces. 
Specifically the motion of kinks is such as to optimize the
position of domains with respect to the random field,
building up a finite response~\cite{corberi2001}. 
The range of times over which
the non-equilibrium pattern is observed can be expanded
by letting $T\to 0$ (keeping $h/T$ small, for linear
response theory to hold), or $J\to \infty $.
We refer to the latter limit for simplicity,
that has the advantage of being easily implemented numerically by
forbidding the flip of spins in the bulk.
With $J=\infty $, $\tau _{eq}^{NCOP}=\infty$ and the system never
equilibrates. The exact solution of the model~\cite{godreche2000}
with $J=\infty $ yields the aging form 
\be
T\chi (t,t_w)= \frac{\sqrt{2}}{\pi} \arctan 
\sqrt{\frac{t}{t_w} -1}
\label{jinfty}
\ee
that converges, in the large $t$ limit, to 
\be
T\chi _\infty=\frac{1}{\sqrt 2}
\label{rad}
\ee
for large $t$.
For finite $J$, as already anticipated, the response of the model
is the same as with $J=\infty $ for times $t\ll \tau _{eq}^{NCOP}$ while,
for larger times, the equilibrium susceptibility is recovered. 

For what concerns the generalization~(\ref{cug}) of the FDT,
notice that~(\ref{jinfty}) obeys the scaling form 
$\chi (t,t_w)=\hat \chi (t_w/t)$. Hence eliminating $t_w/t$
with $C(t,t_w)=\hat C(t_w/t)$ with $J=\infty $ one finds
\be
\chi (C)= \frac{\sqrt{2}}{\pi} \arctan 
\left [\sqrt{2} \cot \left (\frac{\pi}{2}C
\right ) \right ].
\label{cuglia22}
\ee
This curve is plotted in Fig.~\ref{fdtncop}. In the limit
$C(t,t_w)\to 0$, namely $t\to \infty $, the
value $\chi _\infty$ is recovered, as previously discussed.
For finite $J$ the behavior of $C(t,t_w)$ and $\chi (t,t_w)$ 
is unchanged with respect to the case $J=\infty $ up to $\tau _{eq}^{NCOP}$.
Therefore, in the fluctuation-dissipation plot in Fig.~\ref{fdtncop}
the same curve as with $J=\infty $ is followed from $C(t_w,t_w)=1$ down
to $C_{eq}=C(\tau_{eq}^{NCOP},t_w)$.
For $C(t,t_w)<C_{eq}$, namely for $t>\tau _{eq}^{NCOP}$,
the system goes to equilibrium, $\chi (C)$ departs from the master curve 
with $J=\infty$
and approaches $\chi _{eq}$. For fixed $t_w$, $C_{eq}$ grows with
temperature, because $\tau _{eq}^{NCOP}$ decreases as $T$ is increased.
Alternatively, for a given temperature,  
$C_{eq}$ grows by increasing $t_w$.
In conclusion, the master curve is followed in a wider range
by decreasing $t_w$ or $T$, because in this way $C_{eq}$ is reduced.
This explains the behavior of $\chi (C)$ in Fig.~\ref{fdtncop}.

For Kawasaki dynamics the kinetic process is more complex than
in the non-conserved case. The different coarsening mechanism does not
simply change the growth law exponent, but even the two-time quantities
considered here are radically modified with respect to NCOP. 
In particular, as discussed
in the previous Section, the exponent $\lambda =3/2$ differs
from the value~\cite{Bray1989} $\lambda=1$ for Glauber dynamics.
Given these differences the fact that the fluctuation-dissipation plot 
turns out to be the same for NCOP and COP, as will be shown below,
is unexpected and far from being trivial.

The fluctuation-dissipation plot with COP is shown 
in Fig.~\ref{fdtcop}. Curves with different $t_w$ 
collapse on the same master curve for $C(t,t_w)>C_{eq}$.
The collapse of curves with different $t_w$    
proves the validity of~(\ref{cug}) in the scaling regime
even with COP. Moreover, this master curve is the same as with NCOP.
As for NCOP, the collapse occurs
only when both times $t$ and $t_w$ belong to the scaling regime.
With NCOP scaling is obeyed starting from a microscopic
time $t_0$, which is temperature independent\cite{Glauber1963};
with COP this regime is entered after $\tau _{ev}$, which depends on $T$. 
Therefore, for low temperatures
the time scales over which curves with different $t_w$ coincide
are completely different in the two cases, as revealed
from the waiting times reported in Figs.~\ref{fdtncop} and~\ref{fdtcop}.
Comparing the two Figures, one also concludes that changing $t_w$ or
$T$ only produces a shift of $C_{eq}$, the point where $\chi (C)$
deviates from the master curve, in complete analogy for both dynamics.
Then, on the basis of what is known with NCOP, we expect
in the zero temperature limit the  master-curve to be followed down
to $C(t,t_w)=0$.
Let us stress that the correlation and the response
functions {\em are different} for the two dynamics. Only when the response
$\chi(t,t_w)$ is expressed as a function of the correlation $C(t,t_w)$
one obtains {\em the same} fluctuation-dissipation relation $\chi(C)$.

The universality of the non-equilibrium response with respect to
the type of dynamics naturally raises the question of a possible
common fundamental origin.
As proposed in Section~\ref{intro},
this must be of a dynamical character, since the connection~(\ref{statdin})
between statics and $\chi(C)$ cannot be invoked in $d=1$.
The main dynamical feature of phase-ordering is the presence of a
coarsening structure with many competing domains.
This suggests to look for the underlying universality
in the response of a single domain to the perturbation.
In order to test this idea, in the following Sections we
investigate simplified models for the motion of a single domain.
Underlying this approach is the assumption that $\chi (t,t_w)$ of the whole
system can be seen as the sum of the response of single domains
considered independent.
Correlations between domains are only responsible for the growth of their
typical size.
In this way we are able to identify the diffusive wandering of domains
as the origin of the non-equilibrium response, both for NCOP and
COP dynamics.

\section{A single domain model with rigid diffusion} \label{onerigid}

Let us consider an isolated domain ${\cal D}_l$ of up spins 
covering the segment
$[j,j+l]$ of an infinite one-dimensional lattice.
A quenched random variable $h_i$ is defined on the sites of the
lattice via~(\ref{campo},\ref{meanh},\ref{variah}).
At each time step ${\cal D}_l$ is allowed to move rigidly one lattice unit 
on the right or on the left with probability given by
~(\ref{flipprob}), with $\Delta E=h_j-h_{j+l+1}$ or
$\Delta E=h_{j+l}-h_{j-1}$ respectively.
The model can be regarded as an Ising chain with an initial condition
containing $l$ up spins in the interval $[j,j+l]$ in a sea of down spins.
This Ising model is governed by a dynamical rule which conserves
the magnetization and allows only rigid translations of the up domain.
$\Delta E$ is then exactly the
energy gain computed through the Ising Hamiltonian~(\ref{ham})
with $i$ running over the sites occupied by the domain.

\begin{figure}
\centerline{\psfig{figure=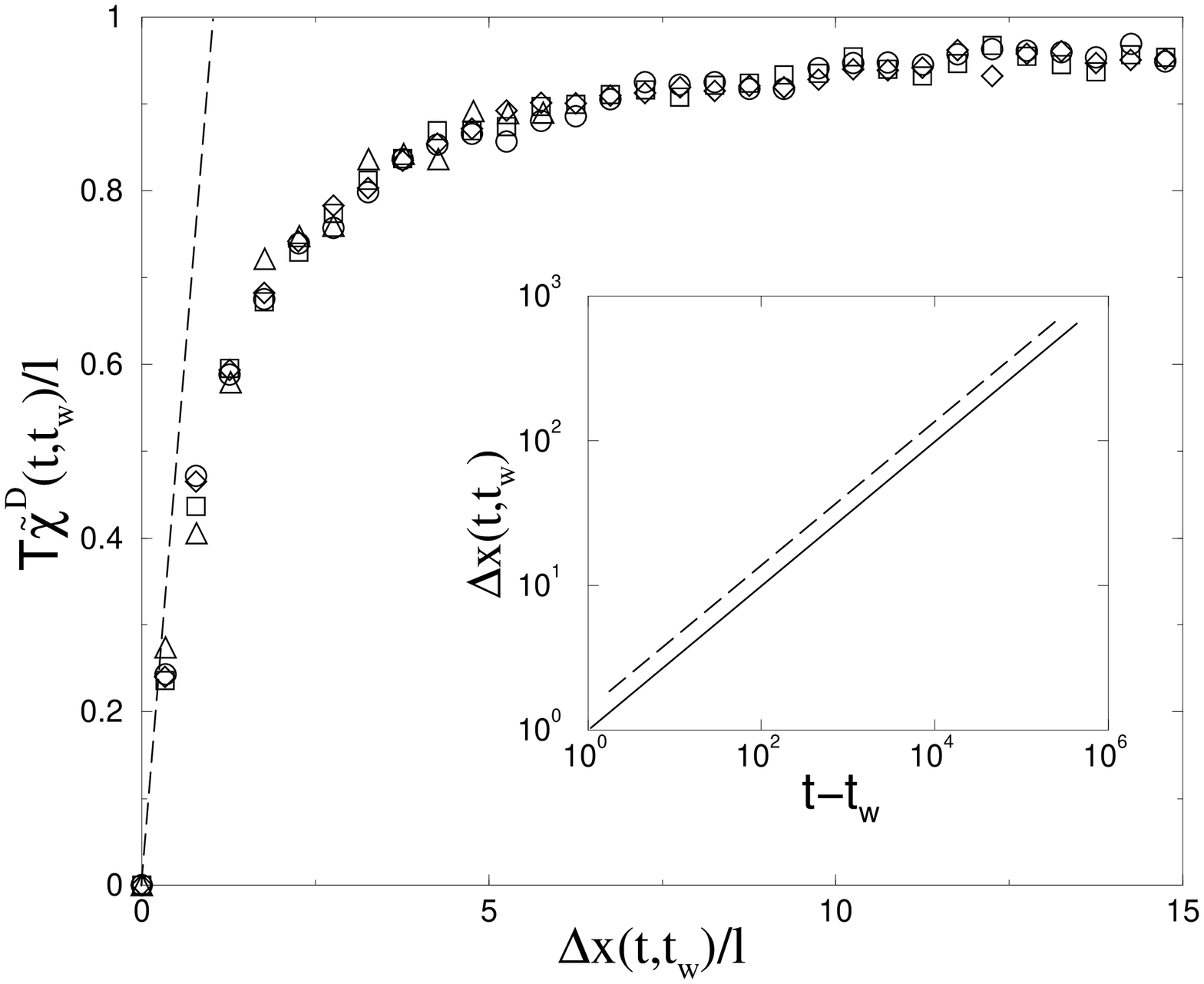,width=9cm,angle=-90}}
\caption{The scaled single domain response 
$T l^{-1}\tilde \chi ^{{\cal D}_l}(t,t_w)$ is plotted against 
$\Delta x(t,t_w)/l$ for $T=0.7$, $h=10^{-2}$
and $t_w=0$. 
Circles, squares and diamonds
correspond to domains of size $l=10,20,40$, respectively. Averages are
taken over $10^6$ trajectories.  
The dashed line is the analytic behavior for small $\Delta x$.
Triangles are the response of a domain whose average size $L(t)$ grows
according to $L(t)=10+\sqrt t/8$. In the inset the behavior of
$\Delta x (t,t_w)$ is plotted for a domain of size $l=40$ with
$t_w=0$.
The dashed line represents the law $(t-t_w)^{1/2}$.}
\label{onedomain}
\end{figure}

For $h=0$ the landscape is flat and the position $x$ of the
center of the domain performs a random walk.
Provided the linear response regime ($h/T\to 0$) is considered,
also for finite $h$ the root mean square displacement 
$\Delta x(t,t_w)$ in a time interval $[t_w,t]$ obeys
$\Delta x(t,t_w)\sim \sqrt {t-t_w}$, as shown in the inset of 
Fig.~\ref{onedomain}.
The response function of the domain is defined as
\be
\tilde \chi ^{{\cal D}_l}(t,t_w)=\lim _{\frac{h}{T}\to 0}
\frac {1}{h} \overline {\langle \sum _ {i\in {\cal D}_l}\epsilon_i \rangle _h}. 
\label{chidom}
\ee
The notation $\langle \cdots \rangle _h$ indicates averaging,
for a single realization of the random field, over the
trajectories of ${\cal D}_l$. 
Notice that, differently from~(\ref{resp}), $i$ runs only over
sites of ${\cal D}_l$. 

Let us consider the behavior of $\tilde \chi ^{{\cal D}_l}(t,t_w)$
for short times.
Initially, at $t=t_w$, the response is zero. By moving one lattice spacing
the sum in~(\ref{chidom}) can change by a value $0,+2,-2$.
In the first two cases the energy is unchanged or decreased;
in the last one $\Delta E>0$ and the probability that such a move is
accepted is $\exp(-2h/T)$.
Then, averaging over the trajectories and the random field,
~(\ref{chidom}) gives
$T\tilde \chi ^{{\cal D}_l}(t_w+1,t_w) = 
\lim _{\frac{h}{T}\to 0}[T/h] \overline
{\langle \sum _ {i\in {\cal D}_l}\epsilon_i \rangle _h}=
\lim _{\frac{h}{T}\to 0}
T[1-\exp (-2h/T)]/(2h)$.
Taking the limit $h/T\to 0$ the linear response
function follows $T\tilde \chi ^{{\cal D}_l}(t_w+1,t_w)=1$.

For long times $\tilde \chi ^{{\cal D}_l}(t,t_w)$ approaches,
assuming equilibration, the static susceptibility 
\be
\tilde \chi ^{{\cal D}_l}_{eq}=\lim _{\frac{h}{T}\to 0}\frac {T}{h}
\frac {\partial \ln Z(h,T)}{\partial h}
\ee
where
\be
Z(h,T)=
\sum _{k=0}^l p_k e^{-\frac {h(l-2k)}{T}}
\label{z}
\ee
is the partition function.
Here $k$ is the number of sites  inside the domain where $\epsilon _i=1$
and $p_k=\ {l \choose k} 2^{-l}$ is the probability
of having a particular value of $k$.
From~(\ref{z}) one easily finds
\be
Z(h,T)=\left [\cosh \left ( \frac {h}{T}\right ) \right ] ^l
\ee
yielding $T\tilde \chi^{{\cal D}_l}_{eq}=\lim _{h/T\to 0}(lT/h)\tanh (h/T)
\left [\cosh \left( h/T \right) \right ]^{l-1} =l$.
 
The behavior of $\tilde \chi ^{{\cal D}_l}(t,t_w)$, obtained numerically,
is plotted in Fig.~\ref{onedomain}, showing the validity of the scaling form
\be
T\tilde \chi ^{{\cal D}_l}(t,t_w) =lg(y)
\label{chising}
\ee
where $y(t,t_w)=\Delta x (t,t_w)/l$.
The scaling function behaves as
\be
g(y)=\left \{ \begin{array}{ll}
	           y	& \mbox{ for $y\ll 1$} \\
		   1	& \mbox{ for $y\gg 1$}
	      \end{array}
     \right .
\label{g}
\ee
in agreement with the analytical results for short and long times.

So far we have studied a model where the size $l$ of the
domain is conserved by the dynamical rule.
However, in order to apply this result to the description of the Ising
chain, where domains coarsen, we consider
now a slightly modified version where the size of the single domain
varies stochastically while growing on average.
In such a situation, we indicate with $l(t)$ the size of the domain
at time $t$, with $L(t)$ the average of $l(t)$, and define $y(t,t_w)$ via
\be
y(t,t_w)=\Delta x(t,t_w)/L(t).
\label{iup}
\ee
We have performed numerical simulations
where the size $l(t)$ of ${\cal D}_l$ was increased with a stochastic rule
such that $L(t)=L(0)+a\sqrt t$, with $L(0)=10$ and $a=1/8$ 
(these values are chosen for numerical convenience). 
Fig.~\ref{onedomain} shows that even in this case the scaling
form~(\ref{chising},\ref{g}) holds.

\subsection{Connection with the Ising model}

From the knowledge of the response of a single domain, we can recover
the behavior of the Ising chain, where many domains compete,
by assuming that the latter
can be adequately described by a collection of quasi-independent
domains of average size $L(t)$. This means that all 
effects produced by correlations between domains, apart from 
the increase of $L(t)$, are supposed not to be relevant.
This assumption will be further discussed in Section~\ref{twokawa}.

The overall response of the Ising model is then given by
\be
\chi (t,t_w)=\sum _{l}{\cal P}(l,t)\tilde \chi ^{{\cal D}_l}(t,t_w)
\label{tuttatutta}
\ee
where ${\cal P}(l,t)$ is the fraction of spins belonging to
domains of size $l$ at time $t$, 
which obeys~\cite{derrida96} the scaling form
${\cal P}(l,t)=L(t)^{-1}f[l/L(t)]$.
The analysis can be carried out more easily with the
approximation $f(x)\simeq \delta (x-1)$, i. e. assuming
that all domains have exactly the same size $L(t)$.
Then
\be
\chi (t,t_w)\simeq
L(t)^{-1}\tilde \chi ^{{\cal D}_L}(t,t_w) \simeq {1 \over T} g(y)
\label{tutta}
\ee
where $y(t,t_w)$ is defined by~(\ref{iup}) and
$\Delta x(t,t_w)$ is the average distance traveled by domains of
size $L(t)$.
On the basis of the large time behavior of $y(t,t_w)$, three situations
can be distinguished, namely $\lim _{t\to \infty }y(t,t_w)=\infty$,
$\lim _{t\to \infty }y(t,t_w)= const. >0$ or
$\lim _{t\to \infty }y(t,t_w)= 0$, giving rise to different values for the
response.

The first case occurs in the equilibrium state of the Ising model because
domains diffuse while their size stays constant
due to the formation of new kinks.
On the basis of~(\ref{tutta}), for the total response of the Ising chain
one obtains $\lim _{t\to \infty}\chi (t,t_w)=\chi _{eq}=1/T$.
This is indeed the value~(\ref{equi}) found for the original Ising chain.

The second case, $\lim _{t\to \infty }y(t,t_w)= const. >0$, occurs
when $\Delta x(t,t_w)$ and $L(t)$ grow with the same exponent.
This happens in the scaling regime, both with spin-flip or Kawasaki
dynamics.
With NCOP this is true because
interfaces are brownian walkers
\be
\Delta x(t,t_w)\sim (t-t_w)^{\frac{1}{2}},
\label{deltancop}
\ee
while it is well known~\cite{Bray1994} that
\be
L(t)\sim t^{\frac {1}{2}}. 
\label{lncop}
\ee

With spin-exchange dynamics the behavior of $\Delta x(t,t_w)$ can be
obtained with the following argument:
The displacement of domains is mediated by the evaporation and condensation
of monomers. When a monomer travels a distance $L(t)$,
leaving one domain and joining the nearest, one of the boundaries 
of each of the two domains is displaced by one lattice unit. 
The number of monomers per unit time that leave a domain and reach the
neighbor instead of recondensing on the original one is 
$\rho(t)$~\cite{cornell91}.
Hence, on average, it takes a time $\sim L(t)$ 
to move a domain by a unitary distance. Since domains move randomly
the law of brownian motion 
\be
\left[\Delta x(t,t_w)\right]^2=2D(t-t_w), 
\label{braw}
\ee
is obeyed, with a diffusion coefficient
\be
D \sim \rho(t).
\label{diffusivity}
\ee
Using the appropriate growth law for COP 
\be
L(t)=\rho (t)^{-1}\sim t^{\frac {1}{3}}
\label{lcop}
\ee
one gets
\be
\Delta x(t,t_w)\sim \left [(t-t_w)/t^{\frac {1}{3}}\right ]^{\frac{1}{2}}.
\label{deltacop}
\ee 
Hence for long times $\Delta x(t,t_w) \sim t^\beta$, with
$\beta =1/2$ for NCOP and $\beta =1/3$ for COP.
Comparing~(\ref{deltancop},\ref{lncop}) and (\ref{lcop},\ref{deltacop})
one concludes, as already anticipated, that 
$\Delta x(t,t_w)\sim L(t)$ regardless of the dynamical rule.
Hence, recalling~(\ref{iup}) one obtains with {\it both} 
dynamics the universal form 
\be
y=y_\infty \left (1-\frac{t_w}{t}\right )^{\frac{1}{2}}
\label{univ}
\ee
where $y_\infty $ is a constant that may depend on the dynamical rule.
Inserting this expression into~(\ref{tutta}) one recognizes that
the global response obeys the scaling behavior $\chi (t,t_w)=\hat \chi (t_w/t)$
which, as discussed in Section~\ref{pert}, is correct for the Ising chain. 
Moreover one also obtains
$\lim _{t\to \infty } y=y_\infty<\infty  $ and, therefore,
an asymptotic non-equilibrium value 
$\chi_\infty =g(y_\infty)/T \neq \chi _{eq}$ is generated.
This limiting value is different from the static susceptibility $\chi _{eq}$
because $y_\infty $ is finite, as opposite to the equilibrium
case discussed above where $\lim _{t\to \infty}y= \infty $.
This is in agreement with the behavior of the original Ising
model~(\ref{rad}), with both dynamics. 

Our approach does not allow the evaluation of $y_\infty$.
However it reproduces the main features of the non-equilibrium response
and offers an insight into what goes on
after a perturbation has been switched on in the one-dimensional Ising model.
In particular, the model points out clearly which is the mechanism whereby the
response is produced. Actually, the domain responds 
to the perturbation by moving so as to
optimize its position with respect to the random field.
Such a sharp statement is made possible in this context by the fact that
the dynamical rules do not allow any other possibility and
indicates that the same mechanism is at
work also in the Ising chain. The discussion presented
insofar shows also that another physical ingredient plays
a fundamental role in $d=1$: The convergence to a finite value of
$y(t,t_w)$, namely $\lim _{t\to \infty }y(t,t_w)=y_\infty\neq 0$.
This property holds because
the displacement of domains in $d=1$ is proportional to their average size.
With these ingredients the one-domain model predicts
a finite limiting value $\lim _{t\to \infty} \chi (t,t_w)=\chi _\infty$
of the response of the Ising chain.
We emphasize that the property 
$\lim _{t\to \infty }y(t,t_w)=y_\infty\neq 0$ is far from being trivial.
Although we restrict the analysis in this paper to one dimension,
we believe this to hold only in the $d=1$ case.
As discussed in Section~\ref{intro}, in one dimension domains 
diffuse in order to lower the magnetic energy in absence of 
the additional force 
produced by surface tension, because kinks are point-like objects. 
In $d>1$, instead, the displacement of an interface is not only ruled
by the random field but is also governed by curvature. This additional
mechanism lowers surface tension and competes with the tendency 
to lower the magnetic energy. The weakening of the drift of domains
limits their motion so that $\Delta x(t,t_w)$ grows more slowly than $L(t)$.
As a result $\lim _{t\to \infty }y(t,t_w)=0$ and the response produced in
this way, from~(\ref{chising}) and (\ref{g}), vanishes.
Therefore we expect the third possible behavior of $y(t,t_w)$ 
introduced above to be realized for $d>1$. 

\section{A single domain model with spin exchange dynamics} \label{twokawa}

In the previous Section we have discussed the diffusion of a rod of
average size $L(t)$.  With this dynamics we have obtained the
formula~(\ref{chising}) for the response. 
The connection with the Ising model was then made possible by
scaling arguments where $\Delta x (t,t_w)$ and $L(t)$ were assumed
{\it a priori} to behave as in the NCOP or COP Ising model. 
The dynamics of the rod however is quite different from the actual 
behavior of the Ising model, where generally domains do not move as a whole. 
Furthermore, while diffusion is the mechanism whereby interfaces move in the
Ising model with NCOP, as stated by~(\ref{deltancop}), 
with COP the law~(\ref{deltacop}) is obeyed, showing a non 
brownian character.   
Despite these shortcomings of the model, we have obtained
a good description of the Ising chain and this suggests that
the actual details of the dynamics are irrelevant.

In this Section we introduce an improved single-domain model which  
takes into account the different kinetics for NCOP and COP. 
In particular, for COP, the evaporation-condensation mechanism
which rules the evolution is taken into account.

We study an Ising chain of size $N$ with periodic boundary conditions
and two interfaces initially located in $x_1=1$, $x_2=N/2+1$.
We consider both spin-flip and spin-exchange dynamics, and require
the initial structure with only two domains of opposite sign to
be preserved at all times.
This condition is guaranteed if the temperature is low enough.
However simulations at very low $T$ would be numerically too demanding,
in particular for COP.
For this reason we implement the aforementioned condition in a different way:
With NCOP we let $J=\infty$ by forbidding flips in the bulk.
For COP we use a modified dynamics where new dimers are not allowed to form.
Specifically, when two monomers meet,
one of the two particles is removed and attached randomly to another kink.
With these rules a configuration with only two domains persists.

Let us focus on one of the two domains, indicated with ${\cal D}$,
with the center located in $(x_1+x_2)/2$.
With COP the system evolves via exchanges of monomers between
the two boundaries of ${\cal D}$.
Since at low $T$ at most one monomer is present
the size $l=x_2-x_1$ of ${\cal D}$ is practically conserved. 
With NCOP, on the other hand, interfaces diffuse independently
and the size changes. We consider the range of times $t\ll (N/2)^2$
so that annihilation events do not happen.
Although the size of domains changes, the average value
$L=\overline {\langle x_2-x_1 \rangle _h }$ is constant, due to symmetry.
Then $L=const.$ with both types of dynamics.
Our goal is to compute 
\be
\chi ^{\cal D}(t,t_w)=\lim _{\frac{h}{T}\to 0}\frac {1}{Nh} 
\sum _ {i=1}^N \overline {\epsilon _i \langle s_i\rangle _h }. 
\ee
Notice that this is exactly the response~(\ref{resp}) defined
for the Ising model. The subscript ${\cal D}$ simply reminds
that we are in a situation with only two domains of fixed size.

Before discussing the behavior of the model let us comment on the 
relationship between this approach and the one presented
in the previous Section. An obvious difference is the presence
of two domains instead of one. However, with the choice
$L=N/2$ the domains are equivalent and this merely doubles
the response.
On the other hand, in the regime $t\ll (N/2)^2$
considered now, $y(t,t_w)=\Delta x(t,t_w)/L \ll1$.
Therefore, if this model is equivalent to the previous one
we expect to recover the results with $y(t,t_w) \ll 1$ of the previous Section.

With NCOP the behavior of the model can be deduced from the knowledge
of the response $\chi _{sing}(t,t_w)$ of the case with a single kink 
located in $x(t)$ and
fixed boundary conditions discussed in~\cite{corberi2001}.
For this system it was shown exactly that
\be
\chi _{sing}(t,t_w)=\lim _{\frac {h}{T}\to 0}\frac {1}{Nh} 
\sum _ {i=1}^N \overline {\langle s_i\rangle _h \epsilon_i} =
\frac{2}{NT}\delta x (t,t_w)\sim (t-t_w)^{\frac{1}{2}}
\label{rs}
\ee
where $\delta x(t,t_w)=\langle \mid x(t)-x(t_w)\mid \rangle$ 
is the average distance traveled by the kink
in the time interval $[t_w,t]$.
Result~(\ref{rs}) allows one to deduce the behavior of the present model.
Denoting by $\chi^{(1)}_{sing}(t,t_w)$ and $\chi^{(2)} _{sing}(t,t_w)$
the responses associated to the two interfaces
the total response is simply given by
\be
\chi ^{\cal D}(t,t_w) 
\sim \chi^{(1)}_{sing}(t,t_w) + \chi^{(2)} _{sing}(t,t_w)
\sim \delta x_1 (t,t_w)+\delta x_2 (t,t_w)
\ee
because the two interfaces are independent.
Indicating with 
$\Delta x(t,t_w)=[\delta x_1(t,t_w) +\delta x_2(t,t_w)]/2\propto (t-t_w)^{1/2}$
the average distance traveled by ${\cal D}$ in the interval $[t_w,t]$
one finds
\be
\chi ^{\cal D}(t,t_w)\sim \Delta x (t,t_w)\sim (t-t_w)^{\frac{1}{2}}.
\label{rsd}
\ee

Going back to the Ising model, assuming again that domains 
are non-interacting, the response 
is obtained by multiplying $\chi ^{\cal D}(t,t_w)$ times
the  number of domains present $\rho (t)\sim t ^{-1/2}$, 
\be
\chi (t,t_w)\simeq \rho (t)\chi ^{\cal D}(t,t_w),
\label{singeff}
\ee
yielding
\be
\chi (t,t_w)= \chi _\infty \left ( 1-\frac {t_w}{t} \right )^{\alpha},
\label{pres}
\ee
where $\alpha =1/2$ and $\chi _\infty$ is the asymptotic value. 
Then, from~(\ref{chising},\ref{g},\ref{univ}) 
one recovers the behavior of the previous model in the 
small $y(t,t_w)$ limit, as expected.

\begin{figure}
\centerline{\psfig{figure=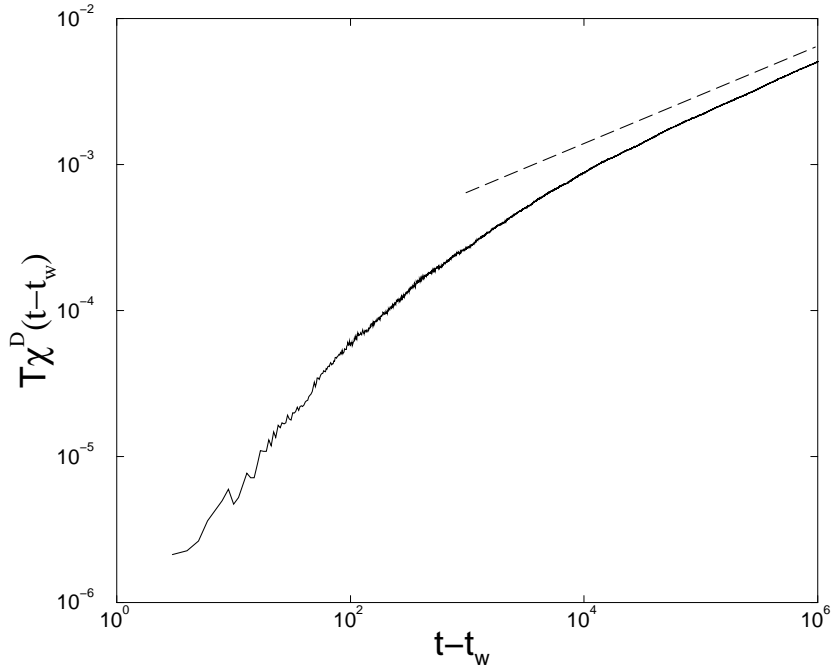,width=9cm,angle=-90}}
\caption{The response of a pair of domains diffusing via
Kawasaki dynamics with probability~(\ref{flipprob2}).
Simulations are presented for a system of $N=10^5$ spins
with $T=0.7$ and $h=10^{-2}$
averaged over $2\cdot 10^5$ realizations.
The dashed line is the $t^{1/3}$ behavior.}
\label{kawasingle}
\end{figure}

With NCOP this result has been obtained by letting the single interface
move as in the original Ising model, namely with the same update rules
for the spins. With COP this issue is more subtle.
In the Ising model the diffusivity of domains depends on
time via~(\ref{diffusivity}). In the present case,
with a fixed size of the domains, $D$ is constant.
To keep this into account, we consider a spin-exchange dynamics generated
by the modified probability
\be
p_T(s_i,s_{i+1})=\min \left [ n^{-\frac{1}{2}}(t)
e^{-\frac {\Delta E}{T}},1\right ],
\label{flipprob2}
\ee 
where $n(t)$ is a counter of evaporation events.
With this rule the diffusivity is proportional to $n^{-1/2}(t) \sim \rho(t)$,
as in the Ising model.
The response of the model, obtained by numerical simulations,
is plotted in Fig.~\ref{kawasingle}, and
for long times
\be
\chi ^{\cal D}(t,t_w)\sim (t-t_w)^{\frac {1}{3}}.
\label{out}
\ee
From~(\ref{out}) the response of the Ising model is obtained  
through~(\ref{singeff}), leading to the same form~(\ref{pres})
as for NCOP, but with $\alpha =1/3$.
This shows that the present model, evolving with two different dynamical
rules with NCOP or COP, gives rise to different responses.
This agrees with the behavior of the Ising model, where both $\chi (t,t_w)$
and $C(t,t_w)$ are different but $\chi(C)$ is the same.

The approach in terms of single domains is based on the
assumption~(\ref{singeff}) of their quasi-independence. We have shown that
this hypothesis allows a description of the Ising kinetics in terms of
scaling arguments and provides a good agreement with
the original model. We are now in a position to substantiate further
the validity of this assumption by checking the accuracy of~(\ref{singeff}).
   
\begin{figure}
\centerline{\psfig{figure=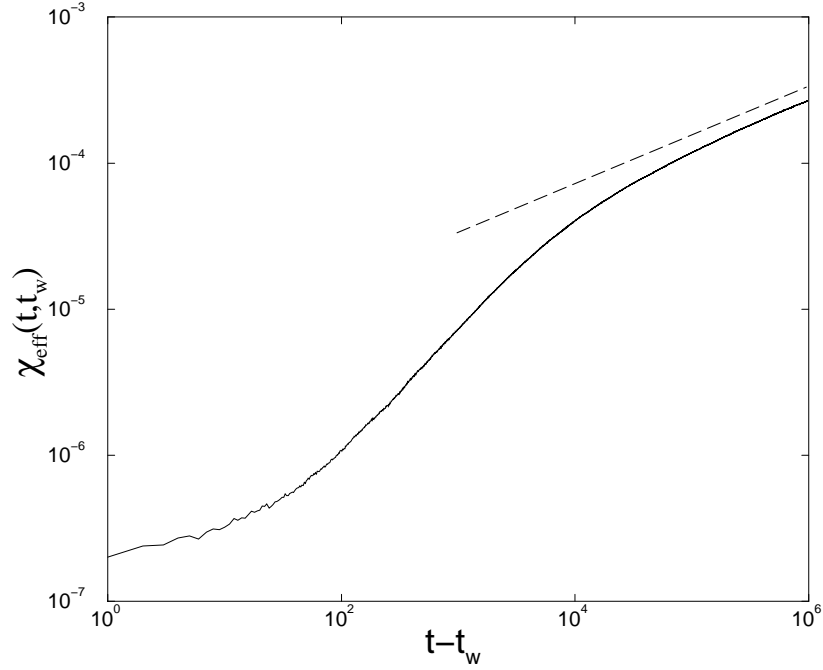,width=9cm,angle=-90}}
\caption{The effective response of the Ising model with Kawasaki
dynamics for a quench to $T=0.48$ with $t_w=3\cdot 10^4$ 
and $h=10^{-2}$.
Data are averaged over 4000 realizations.
The dashed line is the $t^{1/3}$ behavior.}
\label{chieff}
\end{figure}

From the knowledge of $\chi (t,t_w)$ for the Ising model we extract
the effective response due to a single domain defined by
\be
\chi (t,t_w)= \rho (t)\chi _{eff}(t,t_w).
\label{chieffic}
\ee
The accuracy of the independent domain approximation can be determined by
comparing $\chi ^{\cal D}(t,t_w)$ with $\chi _{eff}(t,t_w)$.
With NCOP this issue has been considered in~\cite{corberi2001} showing
a very good agreement. In particular $\chi ^{\cal D}(t,t_w)$ and
$\chi _{eff}(t,t_w)$ both increase as $t^{1/2}$ for large $t$.
For COP the behavior of $\chi _{eff}(t,t_w)$ is shown in Fig.~\ref{chieff}.
For large $t$, in particular, $\chi ^{\cal D}(t,t_w)$ and
$\chi _{eff}(t,t_w)$ grow with the same exponent $1/3$.

The results of this Section point out the robustness of the mechanism
generating the response in $d=1$ which only relies on the coarsening domain
structure of the system. Provided this character is maintained
the global behavior of the susceptibility and, in particular,
the convergence to a finite non-equilibrium value, is the same. 

\section{Discussion} \label{conclu}

In this paper we have studied the off-equilibrium response of the $1d$ 
Ising model. We have shown that the fluctuation dissipation plot
 is the same with NCOP or
COP. In $d=1$, where the connection~(\ref{statdin}) with statics
cannot be invoked,
this universal character has a dynamical origin, as shown by the analysis
of simplified models presented in Sections~\ref{onerigid} and~\ref{twokawa}.

\begin{figure}
\centerline{\psfig{figure=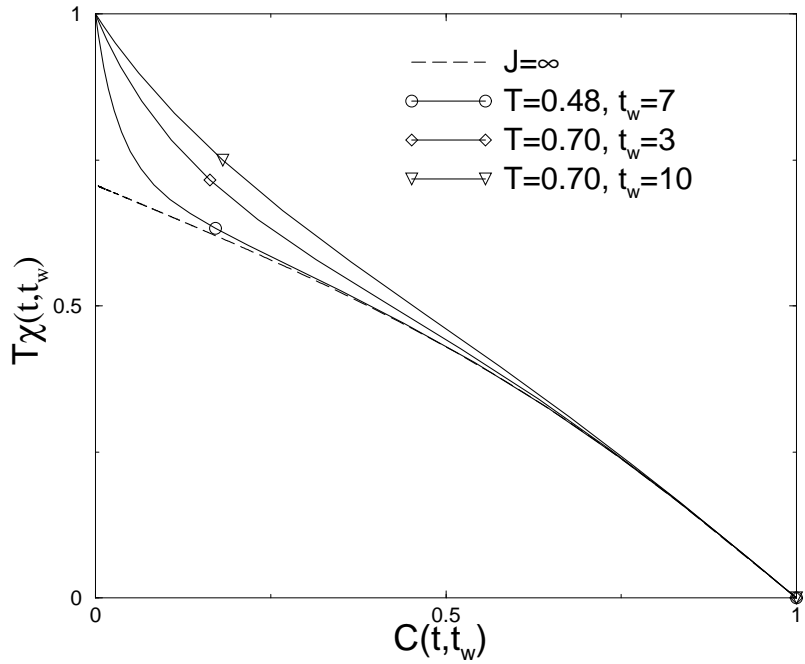,width=9cm,angle=-90}}
\caption{The integrated response function is plotted against
the autocorrelation function with NCOP. 
The dashed line
is the case with $J=\infty $.}
\label{fdtncop}
\end{figure}

\begin{figure}
\centerline{\psfig{figure=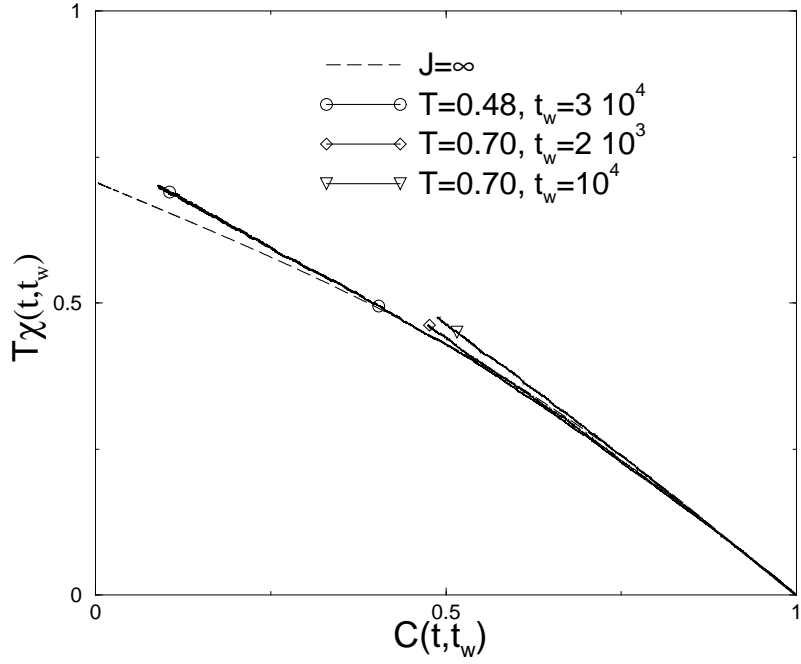,width=9cm,angle=-90}}
\caption{The integrated response function is plotted against
the autocorrelation function with COP. 
Data are obtained from
numerical simulations of a system of $N=10^5$ spins with $h=10^{-2}$, 
averaged over 4000 realizations. 
The dashed line
is the case with NCOP and $J=\infty $, as in Fig.~\ref{fdtncop}.}
\label{fdtcop}
\end{figure}

An important issue is the relevance of the picture provided by the Ising chain
for arbitrary dimension. In $d=1$ with NCOP
the Ising model reaches equilibration on the characteristic time
$\tau _{eq}^{NCOP}=\rho ^{-2}_{eq}=\exp(4J/T)$.
As mentioned in Section~\ref{pert}
this is the time necessary for flipping
spins in the bulk of domains; the same mechanism is also responsible for
the equilibrium response $\chi _{eq}(t-t_w)$.
The off-equilibrium response, instead, develops
in the regime $t<\tau _{eq}^{NCOP}$.
Hence in $d=1$ two kinds of response exist which are observed on different 
time-scales separated by $\tau _{eq}^{NCOP}$.
This feature gives rise to the pattern presented in Fig.~\ref{fdtncop}. 

The case $d>1$ presents some differences.
For quenches below $T_c$ global equilibration is never reached in
an infinite system. Despite this fact the response can still
be split into an equilibrium and an aging part:
The bulk of domains, which behaves as a
pure phase and attains local equilibrium, produces $\chi _{eq}(t-t_w)$ 
which obeys~(\ref{cug}) and~(\ref{statdin}).
Domain walls, instead, are responsible for the non-equilibrium part
which obeys~\cite{corberi2001} the scaling form
\be
T\chi (t,t_w)=t_w^{-a}f\left (\frac{t}{t_w} \right )
\label{scalci}
\ee
with 
\be
a =\left \{ \begin{array}{ll}
                   \frac{d-1}{4} & \mbox{, for $d<3$} \\
                   \frac{1}{2}   & \mbox{, for $d>3$}
	         \end{array}
	\right .
\label{expo}
\ee
and logarithmic corrections in $d=3$. 
The dependence of $a$ on dimensionality 
results from the competition between the drift of interfaces
produced by the perturbation and the force caused by
their curvature. When $a=1/2$, as for $d>3$, the response is simply
proportional to $\rho (t)$ implying that
a single interface produces a response which does not depend on time.
This is what happens if interfacial spins simply {\it polarize}
according to the random field on a microscopic time-scale.
On the other hand, from the knowledge of the behavior of the 
one-dimensional case, we know that the wandering of interfaces
gives rise to a single-interface response growing as $(t-t_w)^{1/2}$.
Therefore a natural interpretation is the following:
Curvature, which is absent in $d=1$  becomes progressively 
more important as $d$ increases, due to the coordination number. 
The attempt to lower surface tension weakens the drift of domain walls
and inhibits the response mechanism associated with it.
This progressively increases the value of $a$ with respect to $d=1$
as dimensionality is increased. Then, for $d>3$ the motion of domain walls 
is fully governed by curvature whereas, for $d<3$
the drift mechanism partly compensate the decrease of $\rho (t)$,
resulting in a smaller exponent $a$.
Only at the lower critical dimension $d=1$, however, 
this mechanism is so efficient as
to balance exactly the loss of interfaces yielding $a=0$ and an
asymptotic finite limit $\chi _\infty$.

For $d>1$ the presence of an equilibrium and an off-equilibrium response,
and the mechanisms whereby they are produced, strongly resemble the
situation in $d=1$. However, while for $d=1$ they are observed on
different time-scales, for $d>1$ they are both developed during the
phase-ordering process. Then, for $d>1$, since the equilibrium part alone
obeys~(\ref{statdin}), in order for the total response to fit into the 
scheme~(\ref{statdin}), the off-equilibrium contribution must vanish
in the large $t_w$ limit. (\ref{expo}) shows that this happens
for $d>1$ but  
the decay of the off-equilibrium response for $d<3$ is slower
than usually expected on the basis of the idea that the random
field simply {\it polarizes } the interfacial spins if $d\leq 3$.

For $d=1$ in the phase-separation regime $\chi _{eq}(t-t_w)$ is
absent and only the off-equilibrium response is developed which
in this case does not vanish for $t_w\to \infty$ and causes
the breakdown of the connection~(\ref{statdin}) with the statics.  

In conclusion, with NCOP an overall discussion of the
response of the Ising model to stochastic perturbations can be given in terms
of two mechanisms whose interplay is regulated by dimensionality.
The equilibrium response, which prevails in $d>1$, 
only relies on the structure of the equilibrium state through~(\ref{statdin}) 
and, therefore, is independent from dynamics. 
In this paper it was shown that in $d=1$ also the off-equilibrium response   
is independent on the kinetic rules,
although this property has a different origin. 
The possible universality of the out of equilibrium response in 
higher dimensionality and the generality of the scaling 
form~(\ref{scalci}) are interesting issues which deserve to be investigated
in the future.

As a final comment, we discuss the possible relevance of our studies
for systems with a vector order parameter with $N$ components. 
In this case spins basically
rotate rather than flip and this is totally different
with respect to scalar systems. The absence of bulk and interfaces
prevents a straightforward extension of the concepts developed in this
paper and the mechanisms by which the response is built up in the vectorial
case is complex and still not well understood. However, 
the exact computation of the response function in the solvable
large-$N$ model~\cite{Corberi2002} has recently shown a pattern which 
resembles the behavior of scalar systems.
Actually in the large-$N$ model the response function can be
explicitly split into an equilibrium and an off-equilibrium part.
It can be shown, moreover, that the former satisfies~(\ref{statdin})
while the latter obeys~(\ref{scalci}) with an exponent
$a$ that vanishes at the lower critical dimension of the model
which, due to the vectorial character, is $d=2$.
This close analogy with the scalar case shows that, although
the microscopic dynamics is different, the same competition between
two mechanisms for the development of the response exists probably for 
every value of $N$ and that
the same scaling relation~(\ref{scalci}) holds.
This suggests the idea that the off-equilibrium response may be
independent from the dynamics also in vectorial systems, 
as we have shown for the Ising model
at the lower critical dimension. 

\dag corberi@na.infn.it \ddag lippiello@sa.infn.it 

\S zannetti@na.infn.it *castella@pil.phys.uniroma1.it

\end{document}